\documentclass[11pt,a4paper]{article}
\usepackage[makeroom]{cancel}
\usepackage{bm}
\usepackage{amsmath}
\usepackage{amsfonts}
\usepackage{amssymb}
\usepackage{graphicx}
\usepackage{authblk}
\usepackage{cite}
\usepackage[left=2.4cm,top=2.4cm,right=2.4cm,bottom=2.4cm,nohead]{geometry}
\DeclareGraphicsExtensions{.png,.jpg,.pdf}
\usepackage{epstopdf}
\usepackage{float}
\usepackage{fouriernc} 
\usepackage{enumitem}
\usepackage[utf8]{inputenc}
\usepackage[english]{babel}
\usepackage{mathtools}
\usepackage{amscd}
\usepackage{mathrsfs}
\usepackage{amsthm}
\usepackage{siunitx}
\usepackage{caption}
\usepackage{subcaption}
\setlist{nolistsep,leftmargin=*}

\usepackage{mhchem} 
\usepackage{cleveref} 
\usepackage{xcolor}
\usepackage{soul}
\usepackage{mdframed}
\usepackage{framed}
\usepackage{nicefrac}
\usepackage{longtable}
\DeclareMathAlphabet{\mathpzc}{OT1}{pzc}{m}{it}

\title{Tutorial on the chemical potential of ions \\ in water and porous materials: \\ transport, isotherms, and electrical double layer theory} 

\makeatletter

\renewcommand\AB@authnote[1]{\textsuperscript{\normalfont#1}}

\makeatother

\author{P.M. Biesheuvel}

\affil{Wetsus, european centre of excellence for sustainable water technology, Leeuwarden, The~Netherlands.}

\date{} 

\newcommand{\s}[1]{\mathrm{_{#1}}}

\newcommand\mydots{\makebox[1em][c]{.\hfil.\hfil.}}

\begin{document}

\maketitle

\begin{abstract}

In this tutorial we discuss the chemical potential of ions in water (i.e., in a salt solution, in an electrolyte phase) and inside (charged) nanoporous materials such as porous membranes. In water treatment, such membranes are often used to selectively remove ions from water by applying pressure (which pushes water through the membrane while most ions are rejected) or by current (which transports ions through the membrane). Chemical equilibrium across a boundary (such as the solution-membrane boundary) is described by an isotherm for neutral molecules, and for ions by an electrical double layer (EDL) model. An EDL model describes concentrations of ions inside a porous material as function of the charge and structure of the material. There are many contributions to the chemical potential of an ion, and we address several of these in this tutorial, including ion volume and the effect of ion-ion Coulombic interactions. We also describe transport and chemical reactions in solution, and how they are affected by Coulombic interactions.
 
\end{abstract} 


\section{Introduction}

The chemical potential of an ion in solution or inside a charged porous material 
is important, 
because changes in the chemical potential across space 
result in transport of ions. 
When chemical equilibrium is reached, for instance between inside and outside a porous material, this is because the chemical potential has become the same, and thus we can calculate ion concentrations inside a material based on the concentration outside. 
This condition of local chemical equilibrium generally holds across a very thin interface where all changes occur in a layer of a few nm thickness, even when there is 
transport of ions across the interface. 
This equilibrium across an interface is an important element in a theoretical description of absorbent materials or to describe transport of ions across a membrane. In this tutorial we 
first describe the chemical potential in general, then describe reactions and transport, next isotherms for the absorption of neutral molecules, and finally we discuss for ions the Donnan balance and the effect of ion volume and ion-ion electrostatic, Coulombic, interactions, which relate to 
the activity coefficient of ions, and osmotic coefficients of aqueous solutions.\footnote{In this tutorial, we alternatingly use the words ion, solute, species, component, and molecule, and they all refer to entities dissolved in a solvent (such as in water). These words do not refer to the solvent itself.}

\section{The chemical potential of an ion}
\label{section_theory}

There are many contributions to the chemical potential of an ion \textit{i} (unit J/mol), either in solution (electrolyte), or inside a porous material, which all add up to
\begin{equation}
\overline{\mu}_i = \overline{\mu}_{\text{ref},i} + \overline{\mu}_{\text{id},i} + \overline{\mu}_{\text{el},i} + \overline{\mu}_{\text{aff},i} + \overline{\mu}_{\text{exc},i} + \overline{\mu}_{\text{cou},i} + \overline{\mu}_{\text{ins},i} + \overline{\mu}_{\text{mm},i} + \overline{\mu}_{\text{centr},i} + \, \dots 
\label{eq_fund_part_0}
\end{equation}
The totality of all these contributions 
is what we call the chemical potential of an ion, $\overline{\mu}_i$. We can also call it simply the potential, or the total (chemical) potential. In some texts, 
chemical potential refers to a subset of the contributions defined above (ideal, 
plus excess, plus Coulombic interactions), but in this tutorial the totality of all terms is the chemical potential. 
The various terms in Eq.~\eqref{eq_fund_part_0} will be discussed below, but they are in the order of Eq.~\eqref{eq_fund_part_0}: a reference term relevant when we 
describe chemical reactions (and in case of adsorption from a solution to a surface), an ideal term which relates to the entropy of ions, 
electrostatic energy when ions are charged, 
an affinity term related to the energy of interaction of an ion with a certain phase or material, an excess term related to volumetric interactions with other molecules and/or the porous material, then the effect of Coulombic interactions of ions with other nearby ions, and a term related to the energy of inserting a molecule against the prevailing pressure. 
Finally we have molecular interactions, i.e., attractions or repulsions 
between molecules, 
and a centrifugal term that is of importance for large molecules such as proteins, viruses, and other colloidal particles which can be separated from a mixture by centrifugation. (At very high centrifugal forces, also heavy ions can be separated.) 

It is a very intriguing and powerful concept that we can add various contributions to the chemical potential of a molecule or ion together, and then gradients in that total potential lead to flow (movement, transport), until all differences in the chemical potential across space are equalized out~\cite{Biesheuvel_2011,Spruijt_2014,Biesheuvel_Dykstra_2020,Biesheuvel_2022}. Flows then cease and we reach 
\textit{chemical equilibrium}. In general, between two positions a few nm apart, chemical equilibrium is always closely approached, even when there is a flow of molecules. 
Interestingly, between those two positions, the individual terms listed in Eq.~\eqref{eq_fund_part_0} can be very different. For instance, the electrostatic term, $\overline{\mu}_{\text{el},i}$, can make a step change across an interface (between two phases that both allow access to molecule \textit{i}), but then one or more of the other terms, for instance the ideal term, $\overline{\mu}_{\text{id},i}$, changes in the opposite direction. In this way the summation of the two terms is the same on both sides of the interface. 

For several terms in Eq.~\eqref{eq_fund_part_0} the related expressions are known, namely
\begin{align} 
\begin{split}
\overline{\mu}_{\text{id},i} & = RT  \ln \left( c_i / c\s{ref} \right)  \\ 
\overline{\mu}_{\text{el},i} & = z_i F V    \\
\overline{\mu}_{\text{ins},i} & = \nu_i P^\text{s}  \\
\overline{\mu}_{\text{mm},i} & = - RT \sum_j a_{i\text{-}j}' c_j 
\label{eq_fund_part_1}
\end{split}
\end{align}
%
%
where in the ideal term, $\overline{\mu}_{\text{id},i}$, $c_i$ is the concentration of a molecule expressed in mol/m\textsuperscript{3}, and $c\s{ref}$ is a reference concentration of $c\s{ref}\!=\! 1\text{~mol/m}^{3}\!=\! 1 \text{~mM}$. Furthermore, \textit{R} is the Gas constant ($R\!=\!8.3144$~J/mol/K), and \textit{T} is temperature in K. The ideal plus reference contributions can be easily derived from the ideal (osmotic) pressure for a system with one component, $\Pi=c_i RT$, when we use the Gibbs-Duhem equation, which relates osmotic pressure to chemical potential, $\partial \Pi / \partial c_i = c_i \partial \overline{\mu}_i / \partial c_i  $ (for only one component in the solvent), and we integrate from $c\s{ref}$ to $c_i$, and from $\overline{\mu}_{\text{ref},i}$ to $\overline{\mu}_i$. The electrostatic energy of an ion, $\overline{\mu}_{\text{el},i}$, is given by the product of the valency, $z_i$, Faraday's constant, \textit{F}, and the electrostatic potential,~\textit{V}, at that position (a mean-field value, averaged on a scale beyond the distances between ions). The pressure insertion term, $\overline{\mu}_{\text{ins},i}$, is given by $\nu_i P^\text{s}$ and depends on the molar volume of an ion, $\nu_i$, and the solution pressure, $P^\text{s}$, which is the hydrostatic pressure, $P^\text{h}$, minus the osmotic pressure, $\Pi$. This term must be included when there is transport, for instance in ion transport through membranes, or during sedimentation or centrifugation of small colloids~\cite{Spruijt_2014}. At chemical equilibrium, this insertion term is constant (i.e., gradients are zero), and can thus be neglected (except when gravity or centrifugation must be included). 
This term can also be neglected when ions are assumed to be infinitely small. A molecular attraction with a strength $a_{i\text{-}j}$ between each couple \textit{i-j} is described by $\overline{\mu}_{\text{mm},i}$. In this evaluation of $\overline{\mu}_{\text{mm},i}$, \textit{j} refers to all types of molecules in a mixture, with \textit{i} one of them. 
The centrifugal force on a colloid depends on the radial velocity, $\omega$, the distance $r$ from the center of rotation and mass $m_i$, according to $m_i \omega^2 r$. The contribution to the chemical potential is obtained as an integration with distance of the negative of this force. For colloidal particles, gravity can be important and then $\omega^2 r$ is replaced by the gravitational constant, $g$. 
The remaining terms, affinity, excess, and Coulombic interactions between ions, will be discussed in the next sections.


We simplify Eq.~\eqref{eq_fund_part_0} by using a chemical potential that is dimensionless, by dividing all $\overline{\mu}$-terms by \textit{RT}, after which we drop the overbar notation, i.e., $\mu = \overline{\mu}/RT$. 
We use the shorthand notation $\ln c_i$ for the ideal term, leaving out mention of the reference concentration $c\s{ref}$, valid as long as we make sure all concentrations are expressed in the unit of mol/m\textsuperscript{3}=mM. When we discuss adsorption on a surface, $c_\s{ref}$ must be included because $c_\s{ref}$ is different in the two phases, even with a different unit. 
Thus we arrive at
\begin{equation}
\mu_i = \mu_{\text{ref},i} +   \ln c_i  + z_i \phi  + \mu_{\text{aff},i} + \mu_{\text{exc},i} + \mu_{\text{cou},i}+\mu_{\text{mm},i} 
\label{eq_fund_part_2}
\end{equation}
where we also implemented $\phi = V / V\s{T}$, with $V\s{T} $ the thermal voltage, given by $V\s{T} = RT / F $ which at room temperature is 25.6 mV. Here, $\phi$ is the nondimensional electrical potential. 
%
Here we left out the centrifugal term and the pressure insertion term; the latter because we assume mechanical equilibrium and no gravity or centrifugation.

\section{Transport of ions because of gradients in potential}

In this section we describe how gradients in the potential of an ion (molecule, solute) lead to transport. First we consider non-isothermal conditions, and only include the reference and ideal contributions to the chemical potential. Thus, based on Eqs.~\eqref{eq_fund_part_0} and~\eqref{eq_fund_part_1}, we have
\begin{equation}
\overline{\mu}_i = \overline{\mu}_{\text{ref},i} + RT \ln \left(c_i / c\s{ref} \right)  \, .
\label{eq_simplified_mu_tot_tr}
\end{equation}
The rate of transport follows from a force balance acting on an ion, or on a mole of ions, which is
\begin{equation}
\mathcal{F}\s{df} + \mathcal{F}\s{fr} = 0 \, .
\label{eq_simplified_mu_tot_tr_bal}
\end{equation}
The first term in Eq.~\eqref{eq_simplified_mu_tot_tr_bal} is the driving force acting on the ions, which is the negative of the gradient in chemical potential, i.e., 
\begin{equation}
\mathcal{F}\s{df} = - \frac{\partial \overline{\mu}_i }{ \partial x}
\label{eq_simplified_mu_tot_tr_bal_df}
\end{equation}
where we assumed there is only transport in a direction \textit{x}. The term $\overline{\mu}_{\text{ref},i}$ is the specific Gibbs energy, and is a function of temperature, \textit{T}, but not of concentration. Combination of Eqs.~\eqref{eq_simplified_mu_tot_tr} and~\eqref{eq_simplified_mu_tot_tr_bal_df} then leads to
\begin{equation}
\mathcal{F}\s{df} = - \frac{\partial \overline{\mu}_i }{ \partial x } = - \frac{\partial \overline{\mu}_{\text{ref},i}}{\partial T} \frac{\partial T_i}{\partial x} - RT \frac{1}{c_i} \frac{\partial c_i}{\partial x} - R \, \ln \left( \frac{c_i }{ c\s{ref}} \right) \frac{\partial T}{\partial x} \, .
\label{eq_simplified_mu_tot_tr_2}
\end{equation}
The second contribution to the force balance of Eq.~\eqref{eq_simplified_mu_tot_tr_bal} is friction of the molecules of type \textit{i} with other types of molecules or phases, such as the water, and with the matrix structure of a porous material. Assuming friction is with one other phase, for instance water, the frictional force is given by
\begin{equation}
\mathcal{F}\s{fr}= - f_{i\text{-w}} \left( v_i -v\s{w} \right)
\label{eq_simplified_mu_tot_tr_3}
\end{equation}
where the friction coefficient between molecule \textit{i} and water is $f_{i\text{-w}}$, and $v_i$ is the velocity of molecule \textit{i}, and $v\s{w}$ that of the water. 
Combination of Eqs.~\eqref{eq_simplified_mu_tot_tr_bal},~\eqref{eq_simplified_mu_tot_tr_2}, and~\eqref{eq_simplified_mu_tot_tr_3}, leads to
\begin{equation}
- RT \cdot \frac{1}{c_i} \cdot \frac{\partial c_i}{\partial x} - RT \cdot S\s{T} \cdot  \frac{\partial T}{\partial x} - f_{i\text{-w}} \cdot \left( v_i -v\s{w} \right) = 0 
\label{eq_simplified_mu_tot_tr_4}
\end{equation}
where we introduced the Soret coefficient (unit $\text{K}^{-1}$), given by
\begin{equation}
S\s{T} = \frac{1}{RT} \, \frac{\partial \overline{\mu}_{\text{ref},i}}{\partial T} + \frac{1}{T} \, \ln \left( \frac{c_i }{ c\s{ref}} \right)  
\end{equation}
which for a salt solution changes sign from negative at a low salt concentration to positive at a high concentration. We implement 
$D_i = RT / f_{i\text{-w}}$, and rewrite Eq.~\eqref{eq_simplified_mu_tot_tr_4} to
\begin{equation}
v_i = v\s{w} - D_i \left( \frac{1}{c_i} \frac{\partial   c_i    }{ \partial x } + S\s{T} \frac{\partial T}{\partial x} \right)
\label{eq_transport_2}
\end{equation}
and because the molar flux $J_i$ (dimension mol/m\textsuperscript{2}/s) is $J_i = c_i v_i$, we arrive at
\begin{equation}
J_i = v\s{w} c_i - D_i  \left( \frac{\partial   c_i  }{ \partial x } + c_i S\s{T}    \frac{\partial T}{\partial x} \right)
\label{eq_transport_2a}
\end{equation}
which describes diffusion because of concentration gradients and temperature gradients.

From this point onward we discuss isothermal conditions, and thus leave out the term based on the temperature gradient, $\partial T / \partial x$. Eq.~\eqref{eq_transport_2a} then simplifies to
\begin{equation}
J_i = c_i v\s{w}  - D_i \frac{\partial  c_i }{ \partial x }
\label{eq_transport_3}
\end{equation}
which is Fick's law, see Eq.~(I) on p.~10 in ref.~\cite{Einstein}, extended with convection, $c_i v\s{w}$. Thus Fick's law follows from analysing the gradient in chemical potential, setting that off against friction with a background medium, assuming no temperature gradients, 
only considering the ideal term, and leaving out convection. The approach we just followed to derive Eq.~\eqref{eq_transport_3} can be extended with additional contributions to the chemical potential, and additional frictional contributions. 

As an example of such an extension, we can implement the electromigration term. Thus on the left of Eq.~\eqref{eq_simplified_mu_tot_tr_4} we add a term $- z_i F \partial V / \partial x$, which then modifies Eq.~\eqref{eq_transport_3} to
\begin{equation}
J_i = c_i v\s{w} - D_i \left( \frac{\partial  c_i }{ \partial x } + z_i c_i \frac{\partial \phi}{\partial x}\right)
\label{eq_transport_4}
\end{equation}
which is the Nernst-Planck equation extended with convection.

What we neglected up to now is that in a porous medium only part of the volume is available for transport, and that transport pathways are tortuous. That is described by the factors porosity and tortuosity, and by distinguishing between interstitial and superficial velocities and fluxes. In most cases an equation such as Eq.~\eqref{eq_transport_4} still results, with fluxes $J_i$ and $v\s{w}$ defined to be superficial (i.e., per unit projected, total, cross-sectional area through which the molecules flow), and concentrations defined per unit volume of pore space. The diffusion coefficient, $D_i$, is then modified to include a correction for porosity and tortuosity~\cite{Biesheuvel_Dykstra_2020}.

The above derivation showed that it is a gradient in chemical potential, of a molecule \textit{i}, that is the driving force for transport of that molecule \textit{i} relative to phases that exert friction on it. It is not the osmotic pressure that is the driving force for diffusion, even though that is argued various times in ref.~\cite{Einstein}. That is not possible because there is only one osmotic pressure in a mixture of \textit{N} solutes, but all \textit{N} solutes are subject to different driving forces, some towards more dilute regions, some towards more concentrated regions, and that can never be described solely by the gradient in osmotic pressure, because that is a single parameter. Though osmotic pressure is not the driving force defining the motion of single solutes, it plays a role in a balance of mechanical forces that describes the flow of fluid as a whole.

Chemical equilibrium for a certain molecule is the condition that all driving forces on a mole of that species add up to zero (and thus all frictional forces also add up to zero). Thus $\partial {\mu}_i / \partial x =0$ (if we only consider the \textit{x}-direction) and thus the chemical potential, ${\mu}_i$, is the same at two nearby positions $x_1$ and $x_2$, i.e., $ \left. {\mu}_i \right|_{x_1} = \left. {\mu}_i \right|_{x_2}$. This analysis can be made with the dimensional $\overline{\mu}_i$ as well as the nondimensional $\mu_i$. 
We reach chemical equilibrium when there is no longer a flux, i.e., the flux $J_i$ has become zero. Of course there are still transfers, exchanges, of molecules between nearby positions, but the flux is zero, where flux refers to a `net,' or total, transport of molecules in one or the other direction. It is this net flux which is symbolized by $J_i$. Chemical equilibrium is also closely approached when there still is a flux, i.e., $J_i \neq 0$, as long as we compare two very nearby positions, only a few nm apart, and this situation is relevant for the study of the interface between solution and a (charged) porous material.


\section{An isotherm for the absorption of neutral molecules}

Next we describe the distribution of a neutral solute, \textit{i}, between solution and a porous absorbent material. The valency (charge number) of neutral molecules is zero, i.e., $z_i \! = \! 0$, so the electrostatic contribution $z_i \phi$ can be neglected. We can also neglect Coulombic interactions, $\mu_{\text{cou},i}$, and Eq.~\eqref{eq_fund_part_2} then becomes
\begin{equation}
\mu_i = \mu_{\text{ref},i} +   \ln c_i  + \mu_{\text{aff},i} + \mu_{\text{exc},i} 
\label{eq_fund_part_3}
\end{equation}
where for the moment we have also left out molecular interactions. We consider two phases, a solution phase, described by the index $\infty$, and an absorbent material, for which we use index m. At chemical equilibrium, the chemical potential of molecule \textit{i} is the same in both phases, and thus, based on $\mu_{\text{m},i} = \mu_{\infty,i} $, we obtain
\begin{equation}
 \ln c_{\text{m},i}  + \mu_{\text{aff},\text{m},i} + \mu_{\text{exc},\text{m},i}  = \ln c_{\infty,i} + \mu_{\text{aff},\infty,i} + \mu_{\text{exc},\infty,i}
\label{eq_fund_part_4}
\end{equation}
which we rewrite to 
\begin{equation}
\ln c_{\text{m},i}  = 
\ln c_{\infty,i} -\left(\mu_{\text{aff},\text{m},i} - \mu_{\text{aff},\infty,i} \right) - \left(\mu_{\text{exc},\text{m},i} - \mu_{\text{exc},\infty,i} \right)\label{eq_fund_part_5}
\end{equation}
and then to
\begin{equation}
\ln c_{\text{m},i}  = \ln c_{\infty,i} -\Delta\mu_{\text{aff},i} - \Delta\mu_{\text{exc},i} 
\label{eq_fund_part_6}
\end{equation}
where each $\Delta$ describes a difference in each particular term between inside the material, and in the outside solution, i.e., `$\Delta = \text{m} - \infty$.' We can now rewrite Eq.~\eqref{eq_fund_part_6} to
\begin{equation}
c_{\text{m},i} = c_{\infty,i} \Phi_{\text{aff},i} \Phi_{\text{exc},i}  
\label{eq_fund_part_7}
\end{equation}
where we introduce two partition coefficients (or, distribution coefficients), one for affinity, and one for volume exclusion (or, excess) effects, given by
\begin{equation}
\Phi_{\text{aff},i} = \exp \left(- \Delta\mu_{\text{aff},i} \right) \hspace{5mm} , \hspace{5mm} \Phi_{\text{exc},i} = \exp \left( - \Delta\mu_{\text{exc},i} \right)
\label{eq_def_Phi_aff_exc}
\end{equation}
which we will discuss in a moment. But first let us explain that Eq.~\eqref{eq_fund_part_7} is an \textit{isotherm}. It is a description of the concentration of a molecule in one phase, as function of that in another phase. This dependence is not necessarily linear, and will also be influenced by other absorbing molecules. The two partition coefficients are based on a difference in the related $\mu$-term between the absorbent material and solution, so we do not necessarily need to know absolute values, but only the difference. This is of particular relevance for the affinity-term. This term relates to a general, chemical, preference for a molecule to be in one phase relative to in another.\footnote{A related term is solubility, \textit{S}, but solubility is used more in the context of single-phase systems, such as a gas or a liquid, and not for the distribution of a molecule between one phase and the liquid that fills the pores of a material.} This partition coefficient is not dependent on concentration, but often strongly dependent on temperature. It also relates to the factor \textit{K} in a typical isotherm, as we show below. For the gas/liquid interface, we have the same isotherm, called Henry equation. When one of the phases becomes more crowded, more concentrated, then to describe the isotherm, the affinity effect is not enough, and a volumetric effect comes into play that limits further absorption. 

Important for a porous material is that a certain volume fraction is taken up by the solid structure, the matrix. With \textit{p} the porosity, i.e., the volume fraction open to solutes and water, the volume fraction of this matrix is $1\!-\!p$. From this point onward, all concentrations of ions and solutes are defined on the basis of this accessible volume, i.e., per unit volume of pores, and not per unit of the total material (which are pores and solid matrix together). 
It is possible to formulate all concentrations per unit total volume, and in some cases that is useful, but it also creates complications that can be avoided by defining all concentrations per unit pore volume. 

In a \textit{Langmuir isotherm}, these volume effects are described by a lattice-approach, with a finite number of adsorption sites, $c\s{max}$, that are either available or occupied. This statistical assumption leads to an excess term given by
\begin{equation}
\mu_{\text{exc},i} = - \ln \left( 1- \frac{1}{c\s{max}} \sum_j c_j \right) = - \ln \left(1- \sum_j \theta_j \right)
\label{eq_mu_exc_lattice}
\end{equation}
where we include the possibility of multiple types of molecules, \textit{j}, absorbing onto the sites, and we introduce $\theta_i$ as the fraction of sites occupied by molecule \textit{i}, $\theta_i=c_{\text{m},i}/c\s{max}$. Instead of a formulation based on $c\s{max}$, it is also possible to replace $c\s{max}$ by $1/\nu$, where $\nu$ is the volume per site (or area per site in case we use this theory for adsorption to a surface). If there is only one type of molecule, \textit{i}, and assume that volume effects in solution can be neglected, we arrive at
\begin{equation}
\frac{\theta_i }{1-\theta_i}= K \cdot c_{\infty,i}  
\label{eq_fund_part_11}
\end{equation}
where \textit{K} is defined as $  K = \Phi_{\text{aff},i} / c\s{max} $, and we used $\mu_{\text{exc},\infty}\!=\!0$. Eq.~\eqref{eq_fund_part_11} is the Langmuir isotherm. We derived Eq.~\eqref{eq_fund_part_11} for absorption in a volume, but we can just as well use it for adsorption onto a surface, in which case $c_{\text{m},i}$ and $c\s{max}$ are surface concentrations. For a low \textit{K} or low $c_{\infty,i}$ (i.e., low occupancy $\theta_i$), the isotherm predicts a linear dependence of $c_{\text{m},i}$ on $c_{\infty,i}$  with a proportionality factor $K c\s{max}$ which is equal to $\Phi_{\text{aff},i}$. In practice, in equilibrium absorption experiments we measure an absorption, $\Gamma_i$, for instance in mg/g (mg of absorbing molecules per gram of absorbent), and we fit data with this Langmuir isotherm (or another isotherm) based on $\Gamma_i=\theta_i \Gamma_{\text{max},i}$. Thus the data are described with the capacity $\Gamma_{\text{max},i}$ and the affinity \textit{K}. And in this procedure it does not matter if molecules were assumed to absorb in a volume, or onto a surface.

We can extend this isotherm with molecular interactions in the porous material (we neglect them in solution), which then results in the \textit{Frumkin isotherm}. Based on Eq.~\eqref{eq_fund_part_1}d, we have for one component $\mu_{\text{mm},i}= - a' c_i = - \chi \theta_i$ ($\chi = a'c\s{max}$), and because this interaction is only in the porous material, we have $\Phi_{\text{mm},i}=e^{\chi \theta_i}$. The right side of Eq.~\eqref{eq_fund_part_7} can now be multiplied with this term, which modifies Eq.~\eqref{eq_fund_part_11} to
\begin{equation}
\frac{\theta_i}{1-\theta_i} = K \cdot c_{\infty,i} \cdot \exp\left(\chi \cdot \theta_i \right)
\label{eq_fund_part_12}
\end{equation}
which predicts that beyond a critical strength of molecular attraction 
there can be phase separation in the porous material (or on the surface), with a more dilute gas-like phase coexisting with a condensed phase. In that case, each $\theta_i$ in Eq.~\eqref{eq_fund_part_12} is that in its `own' phase (either the gaslike phase, or the condensed phase). 

If such a phase separation occurs, then at equilibrium a very interesting situation develops. In a system with a fixed volume of solution, and of the absorbent material, a fraction $x_i$ of that material is in the condensed state, and $1-x_i$ is gaslike. With a total amount of molecules in this system fixed, we have an overall mass balance over solution and the two regions in the material, and we have twice Eq.~\eqref{eq_fund_part_12}. That is three equations, but we have four unknowns: $c_{\infty,i}$, the two $\theta_i$'s, and $x_i$. The final equation is that the pressure in the gaslike and condensed phases is the same, and in the Frumkin model on which Eq.~\eqref{eq_fund_part_12} is based, this equation of state is (Planck, 1908)
\begin{equation}
- \frac{\Pi \, \nu}{RT} =   \ln \left( 1 - \theta \right) + \tfrac{1}{2} \chi \theta^2 \, .
\end{equation}
So, this last equation is evaluated for the gaslike phase and the condensed phase, and the outcome must be the same.

\section{Volume effects and Coulombic interactions between ions }
\label{section_vol_eff_Bjerrum}

Besides the ideal term, the reference term, and affinity, other contributions to the chemical potential of an ion or other molecule are also important, and here we discuss volume effects, and Coulombic interactions between ions. The volume effect is caused by ions taking up space, and that space is no longer available for other ions, and this leads to an increase in the chemical potential of all ions. Also the matrix of a material excludes volume, which likewise impacts $\mu_{\text{exc},i}$ of an ion. This index exc relates to the word excess, but it can also be read as exclusion. For a solution of ions that are all described as spheres of the same size, the Carnahan-Starling (CS) equation is the most accurate approach to describe this effect, and is given by
\begin{equation}
\mu_{\text{exc},i} = \frac{3-\phi}{\left(1-\phi\right)^3} -3  = 8 \, \phi + 15 \, \phi^2 + \, \dots
\label{eq_CS}
\end{equation}
where $\phi$ is the volume fraction taken up by all ions in solution.\footnote{In other sections of this paper, $\phi$ is the electrical potential, and $\phi\s{D}$ is the Donnan potential. The symbol $\phi$ is also used for the osmotic coefficient.} The CS-equation assumes that all (hydrated) ions are spherical and of equal size. There are extensions to include mixtures of spherical particles of unequal sizes, to describe molecules as short strings of connected beads rather than spheres, and to describe ions inside dilute or dense porous structures~\cite{Spruijt_2014,Biesheuvel_Dykstra_2020,Castano_2022}. Because in solution ion volume fractions are generally not more than perhaps a few percent, this excess term is in most cases not very important in solution, unless we have very high concentrations, for instance 0.5~M or more. This situation is very different in a porous material, as we discuss in the next box.

\begin{framed}
\noindent \underline{Ion volume effects in porous materials, and resulting selectivity}. Expressions are available for $\mu_{\text{exc},i}$ for ions (or other molecules) in a porous material that take into account their size, their concentration, and the concentration of other molecules, and how dense is the porous material. In such a description the porous material is characterized by its porosity, and by a characteristic pore dimension, $h\s{p}$, which is the ratio of all pore volume, over the internal surface area. This is the area of contact inside the porous structure between the liquid-filled pores and the matrix material. It is the inverse of a factor $a\s{L}$ which is a liquid-solid specific area, i.e., a contact area per unit liquid phase volume, a relevant parameter in chemical engineering calculations. 

In case the absorbing solutes do not interact with one another, but only with the porous material, and if we assume the solutes to be spherical, then the contribution to the chemical potential is $ \mu_{\text{exc},i} = \text{\textonehalf }\alpha'$ where $\alpha'$ is the size of the solute divided by the specific pore dimension, $h\s{p}$, see Ch.~5 in ref.~\cite{Biesheuvel_Dykstra_2020}. This expression is valid for values of $\alpha'$ between 0 and 1. We then arrive at a contribution to $\Phi_i$ that is independent of the concentration of ions in the pores.

However, there are no simple expressions for the case that we simultaneously have interactions of the ions with the porous material and with each other, in which case $\Phi_i$ becomes a function of ion concentrations. The full equations are presented in Ch.~5 of ref.~\cite{Biesheuvel_Dykstra_2020}. We make a calculation for a porous material with porosity 62 vol\%, and in these pores 10 vol\% of space is occupied by two types of particles, one small, one larger. The larger particles have a size that is 10\% larger than that of the smaller ones. In solution outside the material they have the same concentration. The calculation predicts that inside the pores the concentration of smaller particles is $2.0 \times$ that of the larger particles, i.e., a selectivity towards the smaller particles (only 10\% smaller in size than the larger ones) of a factor 2.0. Thus, these volume exclusion effects can be quite significant. In this calculation, the porous medium is described as a dense array of beads with all beads a size twice that of the small particles. 
\end{framed}

Besides volume effects, Coulombic interactions between ions is also of importance. The origin of this term is as follows. If in solution anions and cations would have paths that are uncorrelated, then averaged over time they are in the vicinity of other ions of the same charge as often as they are in the vicinity of ions of opposite charge. Then the totality of all Coulombic interactions is zero and this term would not exist. However, this is not the case, ions of the same charge repel, so their paths are deflected as much as possible. Anion-cation interactions, however, are very different. Ions of opposite charge pull on each other and thus on average the distance between them is less than between ions of equal charge~\cite{Milner_1912,Milner_1913}. Because the Coulombic interaction energy is more negative at these shorter distances, these distances are more likely. 
If the distribution of these distances would be the same in a dilute solution as in a more concentration solution, all of this would not have any impact. However, ions of opposite charge being very near each other becomes more likely (happens more regularly) when the solution becomes more concentrated. This reduces the energy of the ions (the ions are `stabilized') when salt concentration goes up, and that will lead to the chemical potential of ions going down.  

To describe these forces, 
in 2020 a new theoretical approach was developed based on the idea that for an ion its most nearby counterion is the most important, and it is in a region around it. The size of this region depends on the salt concentration: the higher the salt concentration, the closer by will be the first counterion~\cite{Biesheuvel_2020,Conway_1970}. 
%
%
%
A calculation can be made evaluating for every possible separation of an anion and cation that are nearest-neighbours, the probability of that separation and the Coulombic energy. The average Coulombic energy of this anion-cation pair is then calculated, and based on that the contribution to the chemical potential of the ions determined. We will use here the common notation for this contribution to the chemical potential, $\ln \gamma_\pm$, where $\gamma_\pm$ is the activity coefficient of the ion, with index $\pm$ implying that a mean (average) value is considered. For a 1:1 salt, the final result is the extended Bjerrum equation, given by 
\begin{equation}
\ln \gamma_\pm = \mu\s{cou} =  - b \cdot c_\infty^{1/3} -\tfrac{1}{4} \cdot b^2 \cdot c_\infty^{2/3}  + 6 \cdot b^3 \cdot q  \cdot c_\infty
\label{eq_Bjerrum1}
\end{equation}
where the factor \textit{b} is $b =  \, z^2 \, \lambda\s{B} \, \sqrt[3]{N\s{av}}$, with 
$N\s{av}$ Avogadro's number, $N\s{av}=6.022 \cdot 10^{23}\text{~mol}^{-1}$ and 
$\lambda\s{B}$ the Bjerrum length, $\lambda\s{B}=e^2 / 4 \pi \varepsilon\s{r} \varepsilon_0 k\s{B}T$, which at $T\!=\!25~^{\circ}\text{C}$ is $\lambda\s{B}\! = \! 0.716$~nm ($\varepsilon\s{r}\!=\!78.3$), and thus for a 1:1 salt at room temperature, the numerical value of \textit{b} is $b \!= \! 0.0605\text{~mM}^{-1/3}$. Furthermore, $q$ is a dimensionless number that depends on the ion size. This equation works very well for 1:1 salts. For instance, for NaCl, Eq.~\eqref{eq_Bjerrum1} is accurate up to a salt concentration of 1.5~M using $q=0.19$, and accepting a small error, can be used up to 2.0~M. The factor $\epsilon\s{r} T$ is almost independent of temperature, and thus $\lambda\s{B}$ is as well, and thus $b$ also, and thus the outcome of Eq.~\eqref{eq_Bjerrum1} is almost temperature independent ($\epsilon\s{r} T$ changes by 2\% between 5$^\circ$ and 25$^\circ$C).

If we only use the first term in Eq.~\eqref{eq_Bjerrum1}, we obtain the Bjerrum equation, in which $\ln \gamma_\pm$ is proportional to the cuberoot of $c_\infty$, i.e., 
\begin{equation}
\ln \gamma_\pm =  - b \cdot c_\infty^{1/3} 
\label{eq_Bjerrum2}
\end{equation}
which was used by Bjerrum in 1916 and 1919~\cite{Bjerrum_1916, Bjerrum_1919}. His equation, the same as Eq.~\eqref{eq_Bjerrum2}, has a dependence on $z^2$, on the dielectric constant $\varepsilon$, and on the cube root of salt concentration. If we rewrite Eq.~\eqref{eq_Bjerrum2} to $\partial \,^{10}\!\log\gamma_\pm / \partial  \sqrt[3]{c_{\infty,\text{M}}}$, with $c_{\infty,\text{M}}$ the salt concentration expressed in M, we obtain the factor $-0.259$. For this same factor, based on experiments reported in 1910, Bjerrum derived the value $-0.253$, so these two numbers are within a few percent the same. 

\begin{figure} \centering
\includegraphics[width=0.85\textwidth]{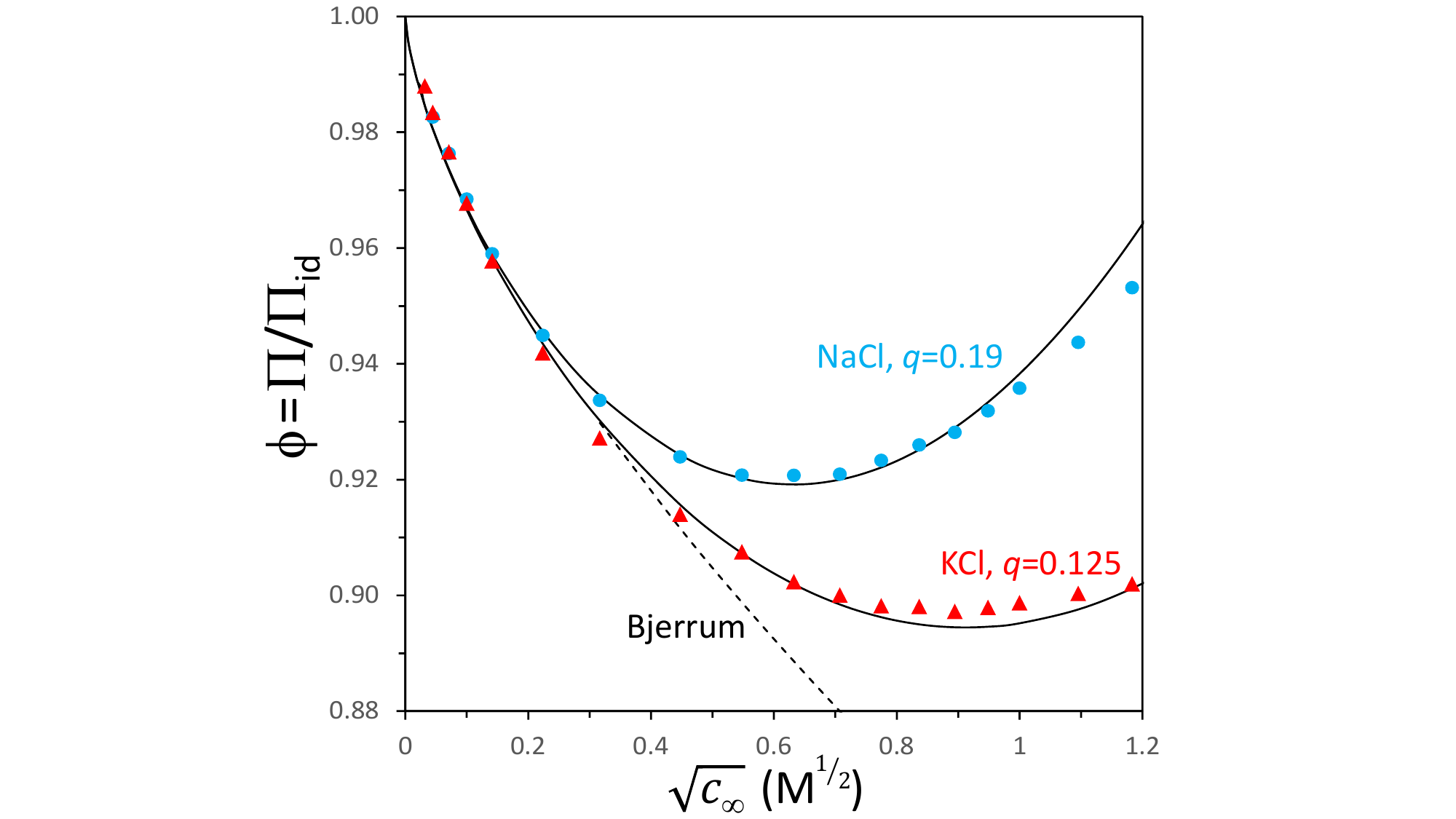}
\vspace{-8pt} \caption{The osmotic coefficient, $\phi$, of a solution of NaCl or KCl as function of salt concentration, $c_\infty$. The osmotic coefficient is $\phi=\Pi/\Pi\s{id}$, with $\Pi$ osmotic pressure and $\Pi\s{id}=2 c_\infty RT$ the ideal osmotic pressure. Solid lines are according to Eq.~\eqref{eq_osm_press_ext_Bjerrum} and points are data. The dashed line is the Bjerrum equation discussed below Eq.~\eqref{eq_osm_press_ext_Bjerrum}.}
\label{fig_osm_coeff}   
\end{figure}

We can recalculate $\ln\gamma_\pm$ to an {osmotic coefficient}, $\phi$, which is the correction to the ideal osmotic pressure because of an activity effect, thus $\phi = \Pi / \Pi\s{id}$. For a $z \!: \! z$ salt, and using Eq.~\eqref{eq_Bjerrum1}, with the ideal osmotic pressure given by $\Pi\s{id} = 2 RT c_\infty$, we find
\begin{equation}
\phi = \frac{\Pi}{\Pi\s{id}} = 1- \tfrac{1}{4} \cdot b \cdot {c_\infty}^{1/3} - \tfrac{1}{10} \cdot b^2 \cdot {c_\infty }^{2/3} +  3 \cdot b^3 \cdot q \cdot {c_\infty }  \, .
\label{eq_osm_press_ext_Bjerrum}
\end{equation} 
In the dilute limit we can leave out the last two terms, which for a 1:1~salt then simplifies to 
\begin{equation}
\phi \sim 1 - 0.151 \cdot \sqrt[3]{c_{\infty,\text{M}}}  \, .
\label{eq_osm_press_simplified_Bjerrum}
\end{equation}
Bjerrum derived the prefactor in Eq.~\eqref{eq_osm_press_simplified_Bjerrum} not of 0.151~$\text{M}^{-1/3}$, 
but `between 0.146 and 0.255 the mean being 0.17'~\cite{Bjerrum_1919}. So again we have a 
close agreement, in this limit of low concentrations. However, this last expression deviates from the exact data beyond $\sim \! 100 $~mM and thus for higher concentrations the full expression, Eq.~\eqref{eq_osm_press_ext_Bjerrum}, must be used. For NaCl and KCl solutions, we evaluate in Fig.~\ref{fig_osm_coeff} data for the osmotic pressure and compare with Eq.~\eqref{eq_osm_press_ext_Bjerrum}~\cite{Biesheuvel_2020}. We use the same values for $q$ as above, which is $q\!=\!0.19$ for \ce{NaCl} and $q\!=\!0.125$ for \ce{KCl}. At 1 M salt concentration, for NaCl, the osmotic coefficient is $\sim \!0.94$, and thus instead of the predicted pressure of $\Pi\sim 50$~bar if an ideal solution would be assumed (based on $\Pi = \Pi\s{id}$), 6\% must be subtracted, and thus the osmotic pressure is $\Pi \sim 47$~bar. If the Bjerrum equation is used, which is the equation just described valid in the dilute limit, the osmotic coefficient is calculated as $\sim \! 0.85$ and then 15\% is subtracted from the ideal value of $\Pi\s{id}=50$~bar, resulting in the prediction that the osmotic pressure is reduced to $\Pi = 42.5$~bar. But this reduction is far too much.

We now return to the activity coefficients of ions, and discuss 2:1 and 3:1 salt solutions. In this case, detailed numerical calculations based on Coulomb's law for more than two ions are required, which are are discussed in ref.~\cite{Biesheuvel_2020}. A very close fit to the data is obtained based on these calculations, but there are no formal analytical solutions that describe these numerical results in a concise manner. Nevertheless, we see in the simulation output and the data, that $\ln \gamma_\pm$ linearly depends on the cube root of salt concentration (i.e., it `scales with the cube root of concentration') until a certain concentration. For \ce{K2SO4}, a 2:1 salt, we have this scaling up to $\sim \! 120$~mM, and thus we can use Eq.~\eqref{eq_Bjerrum2}. We have to use $b\!=\!0.20\text{~mM}^{-1/3}$. For \ce{LaCl3}, a 3:1 salt, we have a linear trend up to $\sim \! 30$~mM, and we can use Eq.~\eqref{eq_Bjerrum2} with $b\!=\!0.30\text{~mM}^{-1/3}$. Thus, for all salt solutions where at least one of the ions is monovalent, a cube root law is followed to beyond 100~mM for 1:1 and 2:1 salts, and up to 30 mM for a 3:1 salt. For the 2:1 salt, we can also use the extended Bjerrum equation, Eq.~\eqref{eq_Bjerrum1}, and have a good fit up to $c_\infty\!=\!0.5$~M based on increasing \textit{b} three times, to $b\!=\!0.1815$~mM\textsuperscript{-1/3} and using $q\! = \! 0.03$. 

For all salts, symmetric and asymmetric, what is now calculated is (ln of) the mean activity coefficient, $\ln \gamma_\pm$. If we assume that for all ions the activity coefficient, $\gamma_i$, is the same, then for all ions we have $\gamma_i= \gamma_\pm$. However, that they are all the same, is an assumption, and without further information, we actually do not know the activity coefficient of an individual ion. Note as well that the above analyses are for a pure salt solution, not mixtures. When all ions are monovalent, then likely we have a reasonably accurate description when we group all ions together as if it is one 1:1 salt, and we use an average value for $q$. But with ions of different valencies, the situation is more complicated and we do not yet have an accurate procedure. 
A first step would be the trace limit of one of the ions, for instance analysis of a single \ce{Ca^2+}-ion in a solution of \ce{NaCl}. An increase in \ce{NaCl}-concentration will lead to a decrease in the chemical potential, $\mu_i$, of the \ce{Ca^2+}-ion, i.e., \ce{Ca^2+} is lowered in energy and the propensity to form of an ion pair involving \ce{Ca^2+} is reduced. This contribution to $\mu_i$ depends on the concentration of \ce{Na+}, not on \ce{Ca^2+}, and thus the terminology of an activity coefficient may be less accurate. 
This contribution of \ce{Na^+} to the chemical potential of \ce{Ca^2+} likely scales with a cube root power, but the correct prefactor is unknown.
 
What these calculations clearly show, is that for fully dissociated salts, the activity coefficient (being different from one) is due to Coulombic interactions between ions, and in addition, at high concentrations there is also the effect of volume exclusion. This knowledge provides us with a good intuition of what to expect when ions are inside a charged nanoporous material. For an uncharged material, ion concentrations are often low (they are excluded, i.e., repelled), 
so for that reason activity coefficients are close to one. 
When a nanoporous material is charged (for instance, a membrane), there is a large concentration of counterions (ions with a sign of charge that is opposite to that of the material) and only a few coions (ions with an equal sign of charge as the material). Because there are mainly counterions, which repel one another, there is no reason to expect activity effects in a similar way as in free solution, where we have similar concentrations of anions and cations, and an attractive anion-cation interaction was driving the decrease of the activity coefficient. So inside a charged nanoporous material, our first intuition is that activity effects because of Coulombic interactions can be neglected. The effects of affinity and volume exclusion are more important. This topic is discussed in more detail in Wang \textit{et al.}~\cite{Wang_2024}.


\section{Reactions between ions in solution}

Ions not only flow or distribute between phases, but can also associate or react with other ions, 
resulting in soluble salt pairs. These salt pairs can grow to colloidal particles, and then form a deposit or sediment~\cite{Kimani_2021}. The equilibrium of a chemical reaction is described by an equilibrium constant, a \textit{K}-value, which depends on temperature and ionic strength. 
At high concentrations and with many types of ions, the complete problem can be quite complex, and is studied in the literature on aquatic chemistry. Special software can be used to calculate the ionic composition of an aqueous solution as function of all ions present, their concentrations, and temperature. 

A special type of reaction is that of ions with hydronium ions, \ce{H3O+}, or in short, with protons, \ce{H+}, such as the protonation reaction of a bicarbonate anion to the neutral carbonic acid. The distribution between \ce{HCO3^-} and \ce{H2CO3} strongly depends on pH. The same is the case for other groups of ions, for instance ammonia/ammonium, or the phosphate system. These protonation reactions can be described without much difficulty based on the pK-values~\cite{Dykstra_2021}. A very different class, and more complicated, is that of redox reactions in water, where one or more electrons are transferred from one species to the other (these molecules can also be from a dissolved gas, or are the solvent itself). Thus, in this reaction where an electron transfer from one species to another, at least two reactant molecules are involved, which both are converted to other molecules. These redox reactions can be slow, but when equilibrium is reached of a redox reaction between all reactants and product molecules, the same theory that we discuss below for ion pairing and protonation, applies as well. 

At chemical equilibrium of a reaction between reactants $1 \mydots n_\s{R}$ and product molecules $n_\s{R+1}\mydots n\s{tot}$ ($n\s{tot}=n_\s{R}+n_\s{P}$), the condition of chemical equilibrium applies, which is
\begin{equation}
\sum_i \nu_i \overline{\mu}_i = 0
\end{equation}
where the summation over \textit{i} includes all reactants and products. The stoichiometric numbers $\nu_i$ are negative for reactants, and positive for products. We include in the chemical potential the reference term, $\overline{\mu}_{\text{ref},i}$, the ideal contribution, $RT \ln \left( c_i / c\s{ref} \right)$, and the Coulombic effects related to the $\ln \gamma_\pm$-terms of the last section~\cite{Wright_2007}. We then arrive at  
\begin{equation}
\sum_i \nu_i \left( \overline{\mu}_{\text{ref},i} + RT \ln \frac{ c_i }{ c\s{ref} } + RT \ln \gamma_i \right) = 0
\label{eq_reaction_3}
\end{equation}
where the factor $\overline{\mu}_{\text{ref},i}$ is the (specific, molar) Gibbs (free) energy, $G_i$. We introduce the activity of an ion, $a_i$, defined as $a_i = c_i \gamma_i / c\s{ref}$, and then rewrite Eq.~\eqref{eq_reaction_3} to
\begin{equation}
\sum_i \nu_i \left( \overline{\mu}_{\text{ref},i} + RT \ln a_i \right) = 0
\label{eq_reaction_3a}
\end{equation}
which can be further rewritten to 
\begin{equation}
K^\text{id}  =  \exp \left\{ - \frac{   \sum_i^{\vphantom{*}} \nu_i \overline{\mu}_{\text{ref},i} }{ RT } \right\}  = \Pi_i^{\vphantom{\nu_i}} a_i^{\nu_i} 
\label{eq_reaction_1}
\end{equation}
where $\Pi_i$ describes the product of activities of all molecules, $1 \mydots n\s{tot}$ (each activity $a_i$ raised to the power $\nu_i$). 
In Eq.~\eqref{eq_reaction_1}, we introduce the ideal reaction equilibrium constant \textit{K}\textsuperscript{id}. For very dilute solutions all activity coefficients are unity, i.e., for all molecules $\gamma_i \! = \! 1$, and then Eq.~\eqref{eq_reaction_1} can be directly solved by replacing all $a_i$'s by $c_i/c\s{ref}$, which leads to
\begin{equation}
K^\text{id}  =  c_\text{ref}^{-\nu\s{tot}}\,\Pi_i^{\vphantom{\nu_i}} c_i^{\nu_i} 
\label{eq_reaction_1a}
\end{equation}
where $\nu\s{tot}=\sum_i \nu_i$.

\begin{figure} \centering
\includegraphics[width=0.58\textwidth]{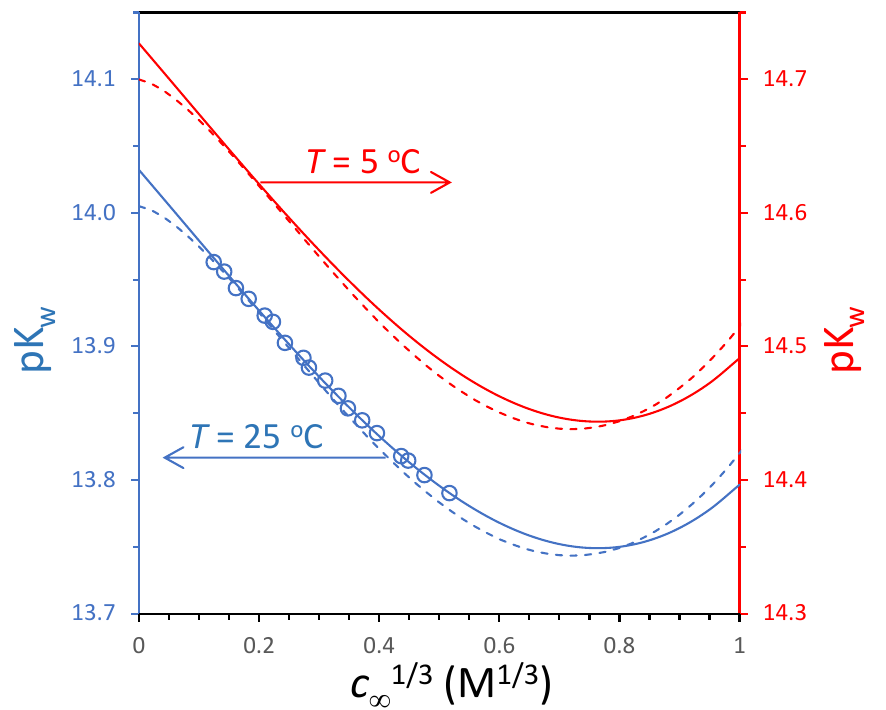}
\vspace{-8pt} \caption{The effect of salt concentration on the water autodissociation constant, pK\textsubscript{w}, at temperatures 5 and 25 $^\circ\text{C}$ based on 
Eq.~\eqref{eq_reaction_7} using $q \! = \! 0.32$ (for $c_\infty\!=\!0$, $\text{pK}\s{w,dil}\!=\! 14.03$ for 25 $^\circ\text{C}$, and $\text{pK}\s{w,dil}\!=\!14.73$ for 5 $^\circ\text{C}$). Results for $T\!=\! 25~^\circ\text{C}$ on left \textit{y}-axis, and for $T\!=\! 5~^\circ\text{C}$ on right \textit{y}-axis. A lower pK\textsubscript{w} implies that \textit{K}\textsubscript{w} is higher, thus concentrations of \ce{OH-} and \ce{H3O+} are higher. Also shown are data from ref.~\cite{Roberts_1930}. Dashed lines are based on the Davies equation used in some commercial software.}
\label{fig_pKw}   
\end{figure}

To illustrate the more detailed analysis that is required when for one or more molecules we have $\gamma_i \neq 1$, we use the example of chemical equilibrium between an anion and cation, A and C, that are formed from two neutral molecules, NM, according to $2 \, \text{NM} \leftrightarrow  \text{A} + \text{C}   $, with  $\nu_\s{A} \! = \! \nu_\s{C} \! = \!  1$, and $\nu_\s{NM} \! = \!   - 2$, thus $\nu\s{tot} \! =\!  0$. Eq.~\eqref{eq_reaction_1} then becomes
\begin{equation}
K^\text{id}=\frac{a_\s{A} \cdot a_\s{C}}{a_{\s{NM}}^2 } \, . 
\label{eq_reaction_4}
\end{equation}
We assume for anion and cation the same activity coefficient (while this is unity for the neutral molecule), and thus arrive at
\begin{equation}
\frac{K^\text{id}}{\gamma_\pm^2}    =  \frac{c_\s{A} \cdot c_\s{C}}{c_{\s{NM}}^2} \, .
\label{eq_reaction_5}
\end{equation}
We now use the extended Bjerrum expression of \S\ref{section_vol_eff_Bjerrum}, Eq.~\eqref{eq_Bjerrum1}, and assume that the neutral molecule, NM, is water, w, and A and C are the \ce{OH-} and \ce{H3O+}-ions, the latter abbreviated as \ce{H+}. We then arrive at 
\begin{equation}
K\s{w}  = K\s{w,dil}  \cdot \left[ \exp \left\{b \, c_\infty^{1/3} + \tfrac{1}{4} \, b^2 \, c_\infty^{2/3} - 6 \, b^3 \, q  \, c_\infty \right\} \right]^2 = [\ce{H+}] [\ce{OH-}] 
\label{eq_reaction_6}
\end{equation}
where $c_\infty$ is the concentration of the 1:1 salt that is added, and $ K\s{w,dil}= K^\text{id} c\s{w}^2 $ (thus, $K\s{w}$ depends on $c_\infty$). Eq.~\eqref{eq_reaction_6} simplifies to 
\begin{equation}
\text{pK}\s{w} = \text{pK}\s{w,dil} - f \cdot \left( b {c_\infty}^{1/3}+ \tfrac{1}{4} b^2 {c_\infty}^{2/3}-6 b^3 q c_\infty \right)
\label{eq_reaction_7}
\end{equation}
where we have $f \!= \! 2\;/\,\ln\left(10\right) \!\sim \! 0.869$.  
We use $q \!= \! 0.32$, which was determined for \ce{HCl} in ref.~\cite{Biesheuvel_2020} to describe the data of $\ln \gamma_\pm$ best. 
We evaluate Eq.~\eqref{eq_reaction_7} in Fig.~\ref{fig_pKw} at two temperatures. 
Because the factor \textit{b} is independent of temperature, the curves are parallel. Also commercial software describing this equilibrium has the same result, with curves of pK$\s{w}$ vs. $c_\infty$ parallel, only shifted up when temperature decreases. We also plot in Fig.~\ref{fig_pKw} calculation output based on the Davies equation that is used in some commercial software, given by $\text{pK}\s{w}=\text{pK}\s{w,dil} - \sqrt{c^*}/\left(1+\sqrt{c^*}\right)-0.315 \, c^*$ with $c^*$ the salt concentration in M. The Davies equation has a scaling in the dilute limit that differs from Eq.~\eqref{eq_reaction_7}, but in the range of available data, the two approaches match rather closely. 

%

Another reaction is $\ce{H2CO3} \leftrightarrow \ce{HCO3-} + \ce{H+}$, and here the equilibrium depends on $c_\infty$ in the same way as in Eq.~\eqref{eq_reaction_6}, with [\ce{H+}][\ce{OH-}] replaced by [\ce{H+}][\ce{HCO3-}]/[\ce{H2CO3}]. However, for a reaction such as $\ce{NH4+} \leftrightarrow \ce{NH3} + \ce{H+}$ the situation is different because there is an ion on both sides, and thus the equilibrium \mbox{$K=[\ce{NH3}][\ce{H+}]/ [\ce{NH4^+}] $} is independent of $c_\infty$ (thus, pK equals $\text{pK}\s{dil}$ at all $c_\infty$).

In Fig.~\ref{fig_pKw}, we can see that $\text{pK}\s{w}$ decreases by 0.7 when the temperature increases from $T\!=\!5\text{~}^\circ$C to $25\text{~}^\circ$C. (This dependence of pK on temperature is based on literature data.) For other pK's, this change is much less. For $\text{pK}\s{dil}$ of \ce{HCO3^-}/\ce{H2CO3}, in that same temperature interval, $\text{pK}\s{dil}$ decreases from 6.517 to 6.351, decreasing further until $\sim \! 50 \! - \! 60\text{~}^\circ$C, and then increasing again (p.~58 in ref.~\cite{Garrels_1960}). 

We make calculations for solutions in equilibrium with either \ce{CO2} or \ce{NH3} from a gas phase (at fixed partial pressure), starting at a condition of no added salt and $[\ce{H+}]\!=\![\ce{OH-}]$, and then we add 1 M of NaCl. We include the activity effect in the calculation. In both cases the result is that the concentrations of ions such as \ce{H2CO3}, \ce{HCO3-}, \ce{NH3}, and \ce{NH4+} are unchanged, and only the concentrations of \ce{H+} or \ce{OH-} change: in the calculation for \ce{CO2} absorption, [\ce{H+}] increases threefold but [\ce{OH-}] is almost unchanged, and in case of \ce{NH3} absorption, [\ce{H+}] is unchanged and [\ce{OH-}] changes by a factor of about 10. Parameter settings re: for water, $\text{pK}\s{w,dil}\!=\!14$; for \ce{CO2} absorption, $\text{pK}\s{dil}\!=\!6.33$ for \ce{H2CO3}/\ce{HCO3^-}; no formation of \ce{CO3^2-}; [\ce{H2CO3}] in solution 1 mM; for \ce{NH3}, $\text{pK}\s{dil}\!=\!9.25$; [\ce{NH3}] in solution 0.1 mM. 

As a final topic, we address the heat associated with such reactions,  for instance of water association/dissociation. This heat is well known as the heat liberated when acid and base are mixed, and when an acid or base is diluted. Here we assume ideal conditions, leaving out the activity effect. 
%
%
So we have an acid solution (volume $V\s{A}$ and a proton concentration $[\ce{H+}]\s{A,0}$), and a base solution ($V\s{B}$ and \ce{OH-}-concentration $[\ce{OH-}]\s{B,0}$). 
After mixing, we have the equality $\left( V\s{A} + V\s{B} \right) \left([\ce{H+}]_\infty-[\ce{OH-}]_\infty \right) \sim V\s{A}[\ce{H+}]\s{A,0} - V\s{B}[\ce{OH-}]\s{B,0}$. This equation can be combined with $K\s{w,dil} = [\ce{H+}]_\infty [\ce{OH-}]_\infty$ and solved. We then know how many water molecules are formed (namely, the total volume after mixing times $[\ce{H+}]_\infty$ minus $V\s{A}$ times $[\ce{H+}]\s{A,0}$). 
With the enthalpy of formation of water from \ce{H+} and \ce{OH-} $-55.8$~kJ/mol at $25^\circ$C, we calculate that the mixing of 1~L of an 1~M acid solution and 1~L of 1~M base solution, leads to the heating of the mixture by $\Delta T = 6.6$~$^\circ$C (heat capacity of water $\rho c\s{p} \! = \! 4.2$~MJ/m\textsuperscript{3}/K). With every factor of 10 lowering of concentration of both solutions, the temperature change goes down by the same factor. If one solution is kept at 1~M, and the other is more and more dilute before mixing, the temperature change after mixing, also drops to zero eventually. Ultimately, when we simply mix an acid such as \ce{HCl} or base such as \ce{NaOH} with pure water, this does not lead to a temperature change from the formation of water.

\section{From Boltzmann's law to chemical equilibrium of a reaction}

\subsection{Lattice gas statistics}

It is interesting to make an analysis based on Boltzmann's famous law for the entropy of a system
\begin{equation}
\textbf{S} = k\s{B} \ln W
\label{eq_Boltz_1}
\end{equation}
where \textbf{{S}} is entropy and \textit{W} is the number of microscopic configurations. 
Here, $k\s{B}$ is the Boltzmann constant, related to the Gas constant, \textit{R}, according to $k\s{B}=R/N\s{av}$. If we assume there are \textit{N} positions for a molecule to be, and \textit{M} molecules, then the number of ways to arrange this, is
\begin{equation}
W = \frac{N !}{M!\left(N \! - \! M\right)! } \, .
\end{equation}
With Stirling's approximation, valid for sufficiently large \textit{N} and \textit{M}, which is \mbox{$\ln N ! \sim N \ln N - N$}, we arrive at
\begin{equation}
\textbf{S} / k\s{B} = 
N \ln N   - M \ln M  - \left(N-M\right)\ln \left(N-M\right) \, .
\end{equation}
We now change from entropy of a system, \textbf{{S}}, to the free energy density, $f = - T \textbf{S} / V$, where \textit{V} is volume. We also use $N = V \cdot c\s{max} \cdot N\s{av} $, $M = V \cdot c \cdot N\s{av}$, and thus $M/N=c/c\s{max}$, and obtain
\begin{equation}
f  = RT \left( c \ln \frac{c}{c\s{max} - c} - c\s{max} \ln \frac{c\s{max}}{c\s{max} - c}  \right)     \, .
\label{eq_Boltz_4a}
\end{equation}
For low values of $c$ relative to $c\s{max}$, Eq.~\eqref{eq_Boltz_4a} simplifies to\footnote{Interestingly, in an expression for \textit{f}, terms that are linear in \textit{c} have no impact on the formulation of $\Pi$ or $\mu_i$ and can therefore always be added or removed. Thus, Eq.~\eqref{eq_Boltz_5} can also be written for instance as $ f = RT c \ln \left(c / c\s{ref} \right) $.}
\begin{equation}
f = RT \left( c \ln \frac{c}{c\s{max}} - c \right) \, .
\label{eq_Boltz_5}
\end{equation}
We can calculate the osmotic pressure, $\Pi$, according to $\Pi = c \, \partial f / \partial c - f$ (assuming only one component), and we then obtain 
\begin{equation}
\Pi = c RT 
\label{eq_Boltz_6}
\end{equation}
which is the ideal gas law, which we here derived from Eq.~\eqref{eq_Boltz_1}. Using the Gibbs-Duhem equation, $\partial \Pi / \partial c_i = c_i \cdot \partial \overline{\mu}_i / \partial c_i$ (in case of only one component), we can integrate from a reference concentration $c\s{ref}$ (where $\overline{\mu}_i = \overline{\mu}_{\text{ref},i}$) to $c_i$, and then obtain from Eq.~\eqref{eq_Boltz_6}
\begin{equation}
\overline{\mu}_i = \overline{\mu}_{\text{ref},i} + RT \int_{c\s{ref}}^{c_i} {c_i}^{-1} \text{d} c_i = \overline{\mu}_{\text{ref},i} + RT \ln \frac{ c_i }{ c\s{ref} }
\label{eq_Boltz_7}
\end{equation}
which are exactly the first two terms of Eq.~\eqref{eq_fund_part_0}, here derived based on Eq.~\eqref{eq_Boltz_1}. 

We can return to Eq.~\eqref{eq_Boltz_4a}, implement a site occupancy, $\theta = c / c\s{max}$, and replace $c\s{max}$ by $1/\nu$, just  as in Eq.~\eqref{eq_mu_exc_lattice}. We then obtain
\begin{equation}
f   = RT \cdot \nu^{-1} \cdot \left( \theta \ln \theta + \left(1-\theta \right) \ln \left(1 - \theta \right) \right)
\label{eq_Boltz_8}
\end{equation}
which appears as if we have an entropy term $\theta \ln \theta$ for the occupied and unoccupied sites added together, though of course there is no physical entity occupying the free sites.

If we evaluate Eq.~\eqref{eq_Boltz_8} to derive the osmotic pressure, we obtain
\begin{equation}
\Pi = - RT \cdot \nu^{-1} \cdot \ln \left( 1-\theta \right)
\label{eq_Boltz_9}
\end{equation}
which in the dilute limit simplifies to the ideal gas law, Eq.~\eqref{eq_Boltz_6}, but at higher concentration increases faster than linearly.

We can also derive an expression for the chemical potential, $\overline{\mu}_i$, based on Eq.~\eqref{eq_Boltz_9}, using again the Gibbs-Duhem equation, and this leads to
\begin{equation}
\overline{\mu}_i = \overline{\mu}^*_{\text{ref},i} + RT \ln \theta_i - RT \ln \left(1-\theta_i \right) 
\label{eq_Boltz_10}
\end{equation}
where $\overline{\mu}^*_{\text{ref},i} = \overline{\mu}_{\text{ref},i} - RT  \ln \left( \theta\s{ref} / \left( 1- \theta\s{ref} \right)\right)$. The last term in Eq.~\eqref{eq_Boltz_10} is the same as the excess term proposed in Eq.~\eqref{eq_mu_exc_lattice} for the Langmuir isotherm, with here a derivation given of that term (for a single component). 

In summary, in this section we developed equations for the ideal gas, and for non-dilute conditions, based on the assumption of a fixed number of positions where a molecule can either reside or not (lattice approach). The ideal gas expression (dilute limit) does not depend on this model but is generally valid: in the dilute limit, what else is possible for osmotic pressure than to be linearly dependent on concentration? However, the non-dilute expressions discussed above, are based on this artificial model of molecules occupying discrete positions. This is likely a good approach when solutes indeed adsorb to distinct surface sites, but to describe the movements of ions in solution, this is not a correct approach, and various expressions relating to the Carnahan-Starling equation for hard sphere mixtures, are much better. We can extend the theory discussed above for a single molecule, to the case of many types of solutes, including also intermolecular interaction energies, such as summarized in Eqs.~\eqref{eq_fund_part_0} and~\eqref{eq_fund_part_1} by the contribution $\overline{\mu}_{\text{mm},i}$.

\begin{framed} 
\noindent \underline{Analysis of Boltzmann's entropy law for ideal molecules}. We can make a simplified calculation where we start with Boltzmann's entropy law, Eq.~\eqref{eq_Boltz_1}, and assume all solutes behave as infinitely small points, i.e., there is no interaction at all between them. A parameter we must use is the number of possibilities to place a molecule in a certain volume, which we give the symbol $\alpha$. So if we assume that in a volume of 1 nm\textsuperscript{3} we can position a certain point-like molecule at a million different positions, then $\alpha\!=\!10^6$~nm\textsuperscript{-3}. We will see that in the analysis we can let $\alpha$ go to infinity, and the outcome will be unchanged (i.e., the end result is independent of $\alpha$).

Thus, we assume we have molecules that behave as ideal points, and they can be placed independent of one another wherever we want in a volume \textit{V}. So each solute can be placed at $\alpha V$ different positions, and thus with \textit{M} molecules, the number of ways to place all \textit{M} molecules in the volume \textit{V}, is $W= \left(\alpha V\right)^M$. Thus the entropy is 
\begin{equation}
\textbf{S}=k\s{B} \ln W = k\s{B} \ln \left(\alpha V\right)^M = k\s{B} M \ln \alpha +  k\s{B} M \ln V \, .
\end{equation}

The free energy of a system is $\textbf{F}=-T \textbf{S}$ and the osmotic pressure is $\Pi = - \partial \textbf{F} / \partial V$, with the differentiation performed at fixed \textit{M}, and thus we obtain for $\Pi$ that
\begin{equation}
\Pi = k\s{B} T M / V = n k\s{B} T 
\end{equation}
where $n=M/V$ is the solute concentration in mol\textsuperscript{-1}. We implement that $n = c \, N\s{av}$ and $k\s{B} = R / N\s{av}$ and thus we arrive at
\begin{equation}
\Pi = c \cdot N\s{av} \cdot {R} \cdot {N\s{av}}^{-1} \cdot T = c RT
\end{equation}
i.e., we derived the ideal gas law from Boltzmann's entropy law without any assumption except that all solutes behave as ideal points that do not interact. The final equation is independent of the parameter $\alpha$ that we introduced earlier on. Importantly, this result follows exactly from Eq.~\eqref{eq_Boltz_1} as long as we assume all solutes have no interaction with one another at all. It does not depend on any other assumption, such as for instance that we are in the dilute limit.

\end{framed}

\subsection{From free energy to chemical equilibrium of a reaction}

It is quite interesting that at chemical equilibrium an equation such as Eq.~\eqref{eq_reaction_1a} describes the relationship between the concentrations of molecules that react with one another. 
Let us demonstrate this again based on minimization of the free energy in a closed system. We consider as an example two molecules, A and B, that can convert from one to the other.  
From integration of Eq.~\eqref{eq_Boltz_7}, we obtain for the free energy density 
\begin{equation}
f =  RT \sum_i \left( c_i  \ln \frac{c_i}{c\s{ref}} - \left( c_i - c\s{ref} \right)\right) + \sum_i \overline{\mu}_{\text{ref},i} \left( c_i - c\s{ref} \right)\,.
\label{eq_free_energy_chem_eq_1}
\end{equation}
This expression for $f$ is different from Eq.~\eqref{eq_Boltz_5} because of the recalculation via osmotic pressure and chemical potential, but in a practical calculation they will lead to the same result. Eq.~\eqref{eq_free_energy_chem_eq_1} is valid for a multicomponent mixture, but only includes ideal entropy and the reference term. The analysis can always be extended with extra contributions. 

If we analyse a closed system with a liquid phase and dissolved molecules A and B, we can find the condition of chemical equilibrium from minimizing the free energy, $f$, with the constraint that matter does not disappear. This implies that $\sum_i a_i c_i$ does not change, where $a_i$ is the elemental number of each molecule, a factor describing how many times a molecule \textit{i} contains an element common to A and B. If we implement this elemental balance in Eq.~\eqref{eq_free_energy_chem_eq_1}, we can write energy $f$ as a function of the concentration of A, and a dependence on B is gone. Because we are at the minimum energy, $\partial f / \partial c_\s{A}=0$ and we arrive at
\begin{equation}
a_\text{A}^{-1} \, \left(\ln \frac{c_\s{A}}{c\s{ref}} + \mu_\s{ref,A}\right) - a_\text{B}^{-1} \,   \left(\ln \frac{c_\s{B}}{c\s{ref}} +\mu_\s{ref,B}   \right)  = 0 \, .
\label{eq_free_energy_chem_eq_3}
\end{equation}

For this problem with two molecules, it is the case that $a_\s{A} \nu_\s{A} + a_\s{B} \nu_\s{B}=0$. We use $a_\s{A} \nu_\s{A} = \alpha$ (for one of the molecules, an arbitrary relation between $a_i$ and $\nu_i$ can be used), and then $a_\s{B} = -\alpha / \nu_\s{B}$, and thus
\begin{equation}
\nu_\s{A} \, \left(\ln \frac{c_\s{A}}{c\s{ref}} + \mu_\s{ref,A}\right) + \nu_\s{B} \,   \left(\ln \frac{c_\s{B}}{c\s{ref}} +\mu_\s{ref,B}   \right)  = 0 
\label{eq_free_energy_chem_eq_4}
\end{equation}
which we rewrite to 
\begin{equation}
\left(\frac{c_\s{A}}{c\s{ref}}\right)^{\nu_\s{A}} \cdot \left(\frac{c_\s{B}}{c\s{ref}}\right)^{\nu_\s{B}} = \exp\left(-\nu_\s{A} \mu_\s{ref,A} - \nu_\s{B} \mu_\s{ref,B}  \right)
\label{eq_free_energy_chem_eq_5}
\end{equation}
which for $\nu_\s{A} \! = \! -1$ and $\nu_\s{B} \! = \! +1$ ($\alpha \! = \! -1$) results in
\begin{equation}
\frac{c_\s{B}}{c_\s{A}} = K = \exp\left(\mu_\s{ref,A} -  \mu_\s{ref,B}  \right)
\label{eq_free_energy_chem_eq_6}
\end{equation}
where $K$ is an equilibrium reaction constant. 
Thus, an equilibrium such as Eq.~\eqref{eq_reaction_1a} also follows from a minimization of free energy. This analysis can be extended to have more molecules and include non-ideal effects, leading to an equation such as Eq.~\eqref{eq_reaction_4}. Thus, a direct analysis based on equality between the chemical potential of all reactants on one side, and products on the other side, is in line with the outcome of a minimization of free energy over all molecules in the system.

\section{Chemical equilibrium involving ions and a charged material}

When a porous material is charged, for instance a membrane used for water desalination, the situation becomes even more interesting. We now assume that in solution, outside the membrane, ions behave ideally, i.e., only the term $\ln c_{\infty,i}$ needs to be considered for the chemical potential. When a porous material is charged, we always have anions and cations, and the simplest case is a 1:1 salt. For an ion absorbing in a charged porous material, we have the effects of affinity and ion volume, just as in Eq.~\eqref{eq_fund_part_7}, and in addition the electrical potential, $\mu_{\text{el},i}=z_i \phi$. The partitioning due to the electrical potential is called the \textit{Donnan} effect, and we then arrive at
\begin{equation}
c_{\text{m},i} = c_{\infty,i} \, \Phi_{\text{aff},i} \, \Phi_{\text{exc},i} \, e^{-z_i \phi\s{D} }
\label{eq_donnan1}
\end{equation}
where the Donnan potential, $\phi\s{D}$, is the electrical potential inside the porous material, $\phi\s{m}$, minus outside, $\phi_\infty$, thus $\phi\s{D} = \phi\s{m} - \phi_\infty$. Just as for the isotherm modeling, we define all concentrations based on the pore volume, i.e., the volume accessible to ions and solvent.

Eq.~\eqref{eq_donnan1} must be set up for both ions, anion and cation, and must be combined with one more equation, for charge neutrality in the membrane, given by
\begin{equation}
\sum_i z_i c_{\text{m},i}  + X = 0
\label{eq_donnan2}
\end{equation}
where \textit{X} is the charge density of the material, expressed as a concentration, i.e., in moles per volume, and just as for ion concentrations, this volume is the water-filled pore fraction. This charge density can be positive (for instance for a membrane for reverse osmosis at low pH) or negative (reverse osmosis membrane at high pH), i.e., it has a sign.  
When $\Phi_{\text{exc},i}$ is a constant (just like $\Phi_{\text{aff},i}$ generally is), i.e., not dependent on ion concentrations, then these equations can be solved jointly without much trouble for any mixture of ions, for a given value of membrane charge, \textit{X}, by inserting Eq.~\eqref{eq_donnan1} for all ions into Eq.~\eqref{eq_donnan2}, and solving for the Donnan potential, $\phi\s{D}$, after which Eq.~\eqref{eq_donnan1} gives us all concentrations in the membrane. This is the same when membrane charge is a function of the Donnan potential. For instance, for a membrane that can charge negatively, 
we can implement in Eq.~\eqref{eq_donnan2} that $X=X\s{max}/\left(1+[\ce{H+}]\s{m}/K\s{A}\right)$, where both $X$ and $X\s{max}$ are negative numbers. The proton concentration in the membrane is also described by Eq.~\eqref{eq_donnan1}, and for protons and hydroxyl ions typically both $\Phi_{\text{exc},i}$ and $\Phi_{\text{aff},i}$ are set to 1, resulting in the Boltzmann relationship
\begin{equation}
c_{\text{m},i} = c_{\infty,i} e^{-z_i \phi\s{D} } \, .
\label{eq_boltzmann}
\end{equation}
Thus, also for an ionizable membrane, i.e., when membrane charge is a function of pH, which in turn is a function of charge and ion concentrations, the resulting equations can be readily solved after iteratively solving for $\phi\s{D}$. 

We now continue with a 1:1 salt as an example, though in principle any mixture of ions can be considered. The expression for charge neutrality becomes
\begin{equation}
c_{\text{m},+} - c_{\text{m},-}  + X = 0 
\label{eq_donnan3}
\end{equation}
and the equation to be solved in $\phi\s{D}$ is
\begin{equation}
c_\infty \Phi_{+}  e^{- \phi\s{D}} -c_\infty  \Phi_{-}   e^{+ \phi\s{D}}  + \frac{X\s{max}}{1+[\ce{H+}]_{\infty} e^{-\phi\s{D} } / K\s{A}} = 0 
\label{eq_donnan_nacl_ionizable1}
\end{equation}
where $\Phi_i$ is the product of the affinity and excess partition coefficients, $\Phi_i = \Phi_{\text{aff},i}\Phi_{\text{exc},i}$, and where  $c_\infty$ is the external salt concentration. 
Now, to simplify, we assume this overall partition coefficient, $\Phi_i$, to be the same for the two ions. We then obtain
\begin{equation}
2 c_\infty \, \Phi_i \, \sinh \phi\s{D}  = \frac{X\s{max}}{1+[\ce{H+}]_{\infty} e^{-\phi\s{D} } / K\s{A}}
\label{eq_donnan_nacl_ionizable2}
\end{equation}
where $\sinh (x) = \text{\textonehalf} \left( \exp (x) - \exp(-x) \right)$. If we now assume that the membrane charge is fixed (for instance because we work at high pH and/or $K\s{A}$ is sufficiently high), we obtain
\begin{equation}
2 c_\infty \,  \Phi_i \, \sinh \phi\s{D}  = X
\label{eq_donnan_nacl_4}
\end{equation}
where we leave out index max. We can rewrite Eq.~\eqref{eq_donnan_nacl_4} to 
\begin{equation}
\phi\s{D}  = \sinh^{-1} \left( \frac{X}{2 \Phi_i c_\infty} \right)
\label{eq_donnan_5}
\end{equation}
where $\sinh^{-1}$ is the inverse of the $\sinh$-function (also written as asinh, or arcsinh), and thus we have an explicit solution for $\phi\s{D}$ as function of salt concentration and membrane charge. The Donnan potential, $\phi\s{D}$, and membrane charge density, \textit{X}, have the same sign. 

Each of the above equations, from Eq.~\eqref{eq_donnan_nacl_ionizable1} to Eq.~\eqref{eq_donnan_5}, is an \textit{extended isotherm}. It is also part of an \textit{electrical double layer} (EDL) theory. An EDL model, or EDL theory, describes the Donnan potential as function of the charge of the material, and based on that, the concentrations of all ions inside a material. Thus an EDL model is an extension of an isotherm, which generally refers to the absorption of neutral molecules. Both isotherms and EDL models describe the relationship between the concentrations of solutes \textit{i} in two phases, when we have chemical equilibrium between these two phases. An EDL model is more extended than an isotherm because it describes solutions with mixtures of ions, not only neutral molecules, and also includes the charge balance for each phase (each phase overall is charge neutral). Thus, the concept of an EDL model refers to the set of equations that describes the structure of a charged interface (or, charged layer) in contact with an electrolyte solution containing ions, including in the description the charge and potential of the interface. 

Based on Eq.~\eqref{eq_donnan_5} we can calculate the anion and cation concentrations inside the charged porous material, again defined per unit pore volume. But first we calculate the total ions concentration, $c\s{T,m}=c\s{m,ct}+c\s{m,co}$, which results in
\begin{equation}
c\s{T,m} = c_\infty \Phi_i \left(e^{\phi\s{D}} + e^{-\phi\s{D}}\right) = 2 c_\infty \Phi_i \, \cosh \phi\s{D}  = \sqrt{X^2+\left(2 \Phi_i c_\infty  \right)^2} 
\label{eq_donnan_6}
\end{equation}
where we used the conversion $\cosh(\sinh^{-1}(x))=\sqrt{x^2+1}$, and implemented Eq.~\eqref{eq_donnan_5}. Eq.~\eqref{eq_donnan_6} shows that the total ions concentration depends on \textit{X} but does not depend on the sign thereof, and is always larger than $|X|$ (the use of $| \mydots |$ refers to the magnitude, i.e., absolute value, of the argument). The concentration of counterions (ct) is larger than $|X|$, and the concentration of coions (co) much smaller. They are given by
\begin{equation}
c\s{m,ct}  = \text{\textonehalf} \left(c\s{T,m} + |X| \right)  \hspace{3mm},\hspace{3mm}
c\s{m,co}  = \text{\textonehalf} \left(c\s{T,m} - |X| \right)  \, .
\label{eq_donnan_7}
\end{equation}
We can check the validity of Eq.~\eqref{eq_donnan_7} by multiplying $c\s{m,ct}$ with $c\s{m,co}$, which should result in $c_\infty^2 \Phi_i^2$, as can be derived fro Eq.~\eqref{eq_donnan1}, which it does. 
Thus, there are counterions in a charged membrane at a concentration larger than the charge density, $|X|$, and coions at a much lower concentration. The concentration of coions is very important as it often determines the salt transport rate in a membrane process. Thus, it is important to exactly know this concentration. Studies of the EDL help in establishing the exact value of this coion concentration. 

Another property of importance is the hydrostatic pressure inside the material, $P^{\text{h}}$, that pushes the charged material outward. The difference in hydrostatic pressure between inside a material and outside, $\Delta P^\text{h}$, equals the change in osmotic pressure between inside and outside, $\Delta \Pi$, and for an ideal 1:1 salt, this difference is $\Delta\Pi = RT \cdot \left( c\s{T,m} - 2 c_\infty \right)$, which can be solved based on Eq.~\eqref{eq_donnan_6}, and then we also know $\Delta P^\text{h}$. For a low external hydrostatic pressure, low charge \textit{X}, and low $\Phi_i$, it is possible that negative pressures develop inside the porous material.

From Eq.~\eqref{eq_donnan3} onward we considered a solution with one anion and one cation inside a charged porous material. In Ch.~2 of ref.~\cite{Biesheuvel_Dykstra_2020} a related problem is described where a monovalent cation, \ce{Na+}, and a divalent cation, \ce{Ca^2+}, absorb from a mixed electrolyte solution into a negatively charged porous material. Observations are, and theory 
predicts this as well, that when the external solution is diluted, thereby keeping the ratio of \ce{Ca^2+} over \ce{Na+} in solution constant, that monovalent cations \ce{Na+} desorb from the material, while divalent cations absorb more. This is somewhat counter-intuitive because normally when a solution is diluted, the absorbed amount decreases. But this is different for a charged material and this phenomenon can be described quantitatively using the balances presented above, without the need to include partition coefficients for affinity or volume. In Eq.~\eqref{eq_donnan_nacl_ionizable1} we only have to include the extra divalent cation. The membrane charge can be set to a fixed value.  

 

\section{Conclusions}

The chemical potential of an ion is an important concept to understand absorption of ions from an electrolyte solution into neutral and charged materials such as absorbents and membranes. This is relevant for processes where ion absorption, transport, and exchange play a key role. For neutral molecules and neutral materials, at chemical equilibrium the concept of an isotherm is of importance, which for charged molecules and materials is extended to an electrical double layer theory. The chemical potential has many contributions, and in a calculation of the distribution of ions across an interface, most terms depend on concentration and temperature. For ions in solution, deviations from ideal behaviour are often classified as activity coefficients, and they are due to Coulombic interactions between ions and volume exclusion. For most conditions, these contributions lead to an osmotic coefficient less than one, and then the osmotic pressure of a solution is less than predicted by the ideal contribution only. In a porous material, the volume of solutes also contributes to the chemical potential, making it more difficult for larger ions to reside in a porous material than smaller ones, and thus ions can be separated based on size.

\end{document}